# A microscopic computational simulation of [$^{18}$F]FDG transport and metabolism identifies valid regimes for compartmental analysis

Xiaoxu Zhong, Guillem Pratx*

School of Medicine, Department of Radiation Oncology and Medical Physics, Stanford University, Stanford, CA 94305, USA

## Abstract

$^{18}$F-fluorodeoxyglucose ([$^{18}$F]FDG) positron emission tomography (PET), combined with compartmental modeling, is a powerful non-invasive imaging method for assessing cellular metabolism. However, classical two- and three-tissue compartment models assume homogeneous [$^{18}$F]FDG distribution within the tissue, which needs justification, and the definition and interpretation of rate constants across these models is not always consistent. To address these issues, we develop a finite difference solver to simulate [$^{18}$F]FDG transport and metabolism within a 1 mm$^3$ tissue volume, representing the smallest volume resolvable by PET. Our simulations reveal sub-millimeter heterogeneity in [$^{18}$F]FDG distribution and show that the measured PET signal is dependent not only on cellular metabolic activity but also on interstitial [$^{18}$F]FDG diffusivity, vascular permeability, and vascular architecture. We further demonstrate that our finite-difference simulation reduces to a three-tissue compartment model when interstitial [$^{18}$F]FDG concentration is homogeneous. Furthermore, this simplified model itself reduces to the two-tissue compartment model when vascular permeability is sufficiently high. This work quantitatively links vascular permeability, vascular architecture, cellular uptake kinetics, [$^{18}$F]FDG diffusivity, and acquisition time. It also unifies the two- and three-tissue compartment models and identifies their applicable regimes. These findings deepen our understanding of [$^{18}$F]FDG transport kinetics and enhance the interpretability of dynamic [$^{18}$F]FDG-PET imaging.

## 1. Introduction

Positron emission tomography (PET) is a metabolic imaging technique that captures cellular-level activity (Zhong et al., 2025). The most commonly used PET radiotracer, $^{18}$F-fluorodeoxyglucose ([$^{18}$F]FDG), is a glucose analog. Following intravenous administration, [$^{18}$F]FDG accumulates in metabolically active cells, causing tumors to appear as regions of high radioactivity (Zhu et al., 2011). This metabolic contrast underpins the clinical utility of [$^{18}$F]FDG-PET in oncology.

Dynamic [$^{18}$F]FDG-PET imaging is often combined with compartmental models (Figure 1) to infer kinetic rate parameters, which are assumed to reflect the underlying transport and phosphorylation of glucose by cells (Kotasidis et al., 2014; Quon and Gambhir, 2005). Although widely used (Bertoldo et al., 2001; Dimitrakopoulou-Strauss et al., 2021; Morris et al., 2004; Pantel et al., 2022), compartment models require assumptions that are challenging to verify experimentally. A prime example is the assumption that [$^{18}$F]FDG distributes homogenously within each compartment, which can fail when limited diffusivity causes spatial gradients in tracer concentration in biological fluids and dense tissue microenvironments. Moreover, compartmental modeling simplifies biological complexity into a set of ordinary differential equations that describe the transport of [$^{18}$F]FDG between connected compartments. However, there are no standardized guidelines for selecting the number of compartments, and the biological interpretation of rate coefficients varies across different compartment modeling approaches.

*Corresponding author
Email address: pratx@stanford.edu

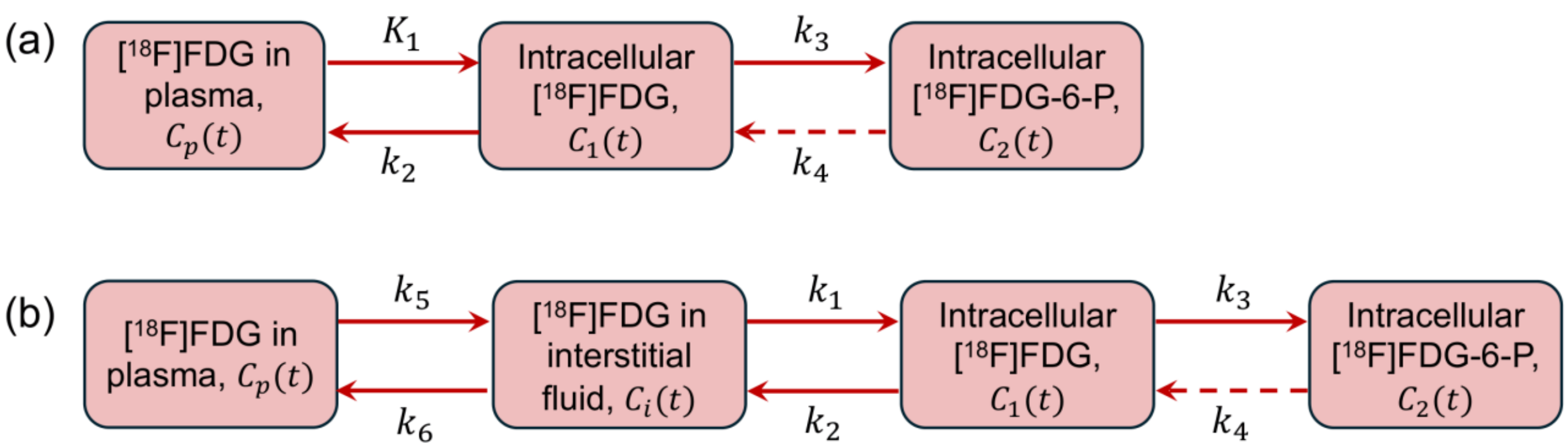

Figure 1. Classical (a) two-tissue compartment model and (b) three-tissue compartment model. Rate constants $k_1$, $k_2$, …, $k_6$ have units of per minute ($min^{-1}$); $K_1$ is in units of milliliter plasma per minute per milliliter tissue ($mL\times min^{-1}\times mL^{-1}$) (Bertoldo et al., 2001; Dimitrakopoulou-Strauss et al., 2021; Kotasidis et al., 2014).

Several studies have investigated [$^{18}$F]FDG transport in the interstitium of tumors and examined the spatiotemporal distribution of [$^{18}$F]FDG (Kashkooli et al., 2022; Shahvandi et al., 2022). However, these works did not address the influence of vascular architecture and [$^{18}$F]FDG diffusivity on [$^{18}$F]FDG-PET interpretation. Additionally, the applicable regimes and the relationships between two- and three-tissue compartment models have not been fully established for complex *in vivo* systems.

This paper aims to address the following two questions concerning compartmental modeling of dynamic [$^{18}$F]FDG-PET data: (1) Is [$^{18}$F]FDG distribution within each voxel of a PET image (~ 1 $mm^3$) spatially uniform? (2) Which factors, other than glucose metabolism, affect voxel intensity (average radioactivity) in PET images? To answer these, we implement a finite-difference computational model simulating [$^{18}$F]FDG transport from blood vessels and its uptake by cells. The model leverages a public dataset (Bumgarner and Nelson, 2022) containing all vascular structures within the inferior colliculus of a mouse, representing a simulation volume of 1 $mm^3$. We observe that [$^{18}$F]FDG distribution becomes more heterogeneous with lower [$^{18}$F]FDG diffusivity in the interstitium, higher cellular uptake rates, and greater cell-vessel distance. We propose two dimensionless parameters that can be used to evaluate the heterogeneity of [$^{18}$F]FDG distribution within tissues and the [$^{18}$F]FDG concentration gradient between interstitium and plasma. This work also clarifies the relationship between the two- and three-tissue compartment models, ensuring consistent interpretation of rate constants across models.

## 2. Methods

### 2.1 Modeling

#### *2.1.1 Finite Difference Model*

The exchange of [$^{18}$F]FDG between the blood plasma, interstitial fluid, and cells is considered in our model (Figure 2(a)). We assume homogeneous [$^{18}$F]FDG concentration in the plasma, given the short blood circulation time, which is typically around 20 seconds. Cell density is assumed to be uniform.

We use $C_p(t)$ and $C_i(t, \boldsymbol{x})$ to represent the [18F]FDG concentrations in the plasma and interstitial fluid, respectively, where $t$ denotes the elapsed time and $\boldsymbol{x} = (x, y, z)$ is the spatial coordinate. The volumes occupied by plasma, interstitial fluid, and cells are denoted as $V_p$, $V_i$, and $V_c$, respectively. Following uptake by cells via glucose transporters, [18F]FDG is phosphorylated by hexokinase enzymes to [18F]FDG-6-phosphate ([18F]FDG-6-P). The intracellular concentration of [18F]FDG and [18F]FDG-6-P, denoted $C_1(t, \boldsymbol{x})$ and $C_2(t, \boldsymbol{x})$, respectively, satisfy the following equations (Dimitrakopoulou-Strauss et al., 2021; Zhong et al., 2025):

$$\frac{\partial C_1}{\partial t} = k_1 C_i - (k_2 + k_3) C_1 + k_4 C_2, \quad (1)$$

$$\frac{\partial C_2}{\partial t} = k_3 C_1 - k_4 C_2, \quad (2)$$

where $k_1$ is the rate constant for [18F]FDG influx into cells, $k_2$ the efflux rate to the interstitial fluid, $k_3$ the phosphorylation rate, and $k_4$ the dephosphorylation rate.

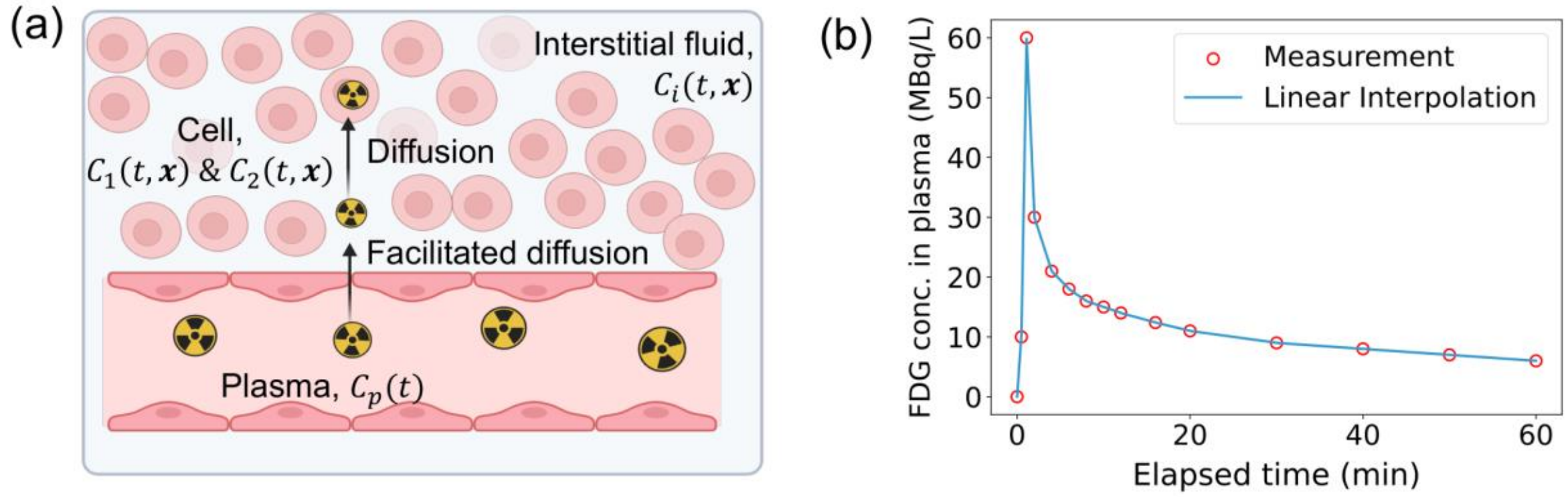

Figure 2. (a) Schematic showing blood vessel, cells, and interstitial fluid distributions. (b) [18F]FDG concentration in the plasma, which is used as a known input function for modeling [18F]FDG kinetics.

To derive the governing equations for the 3D transport of [18F]FDG through the interstitium, we consider a control volume ($v$) consisting of a uniform mixture of interstitial fluid and cells (Figure 3(a)). The volume of the interstitial fluid within this control volume is $v_i = \phi v$, where $\phi = V_i/(V_i + V_c)$. The change of total interstitial [18F]FDG in the control volume is $\partial(C_i v_i)/\partial t = \partial(C_i \phi v)/\partial t$.

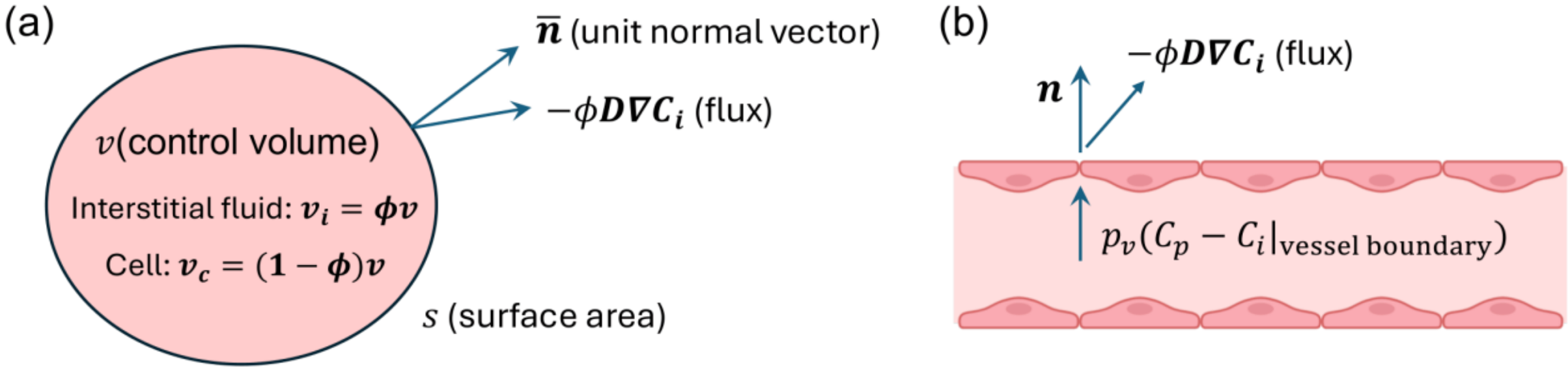

Figure 3. (a) Schematic showing the control volume consisting of a homogeneous mixture of cells and interstitial fluid. (b) Schematic showing the continuous flux of [18F]FDG at the vessel boundaries.

According to the Delesse principle (Oleschko, 1998), the expected area fraction of a phase in a homogeneous mixture is equal to its volume fraction. Therefore, the net inflow rate of [18F]FDG across the surface of the control volume due to diffusion is given by $\int \phi(D\nabla C_i) \cdot \bar{\boldsymbol{n}} ds$, where $\bar{\boldsymbol{n}}$ is the unit vector normal to the control surface, and $D$ is the [18F]FDG diffusivity within the interstitium. Applying the divergence theorem, this net inflow rate of [18F]FDG becomes $\phi D v \nabla^2 C_i$, where $\nabla^2$ is the Laplace operator satisfying $\nabla^2 C_i = \partial^2 C_i/\partial x^2 + \partial^2 C_i/\partial y^2 + \partial^2 C_i/\partial z^2$; here, $x$, $y$, and $z$ are the coordinates of the Cartesian coordinate system.

From equation (1), the net rate of [18F]FDG uptake by cells within the control volume is $v_c(k_1 C_i - k_2 C_1) = (1-\phi)v(k_1 C_i - k_2 C_1)$, where $v_c = (1-\phi)v$ denotes the volume occupied by cells within the control volume. Applying the principle of mass conservation to the interstitial [18F]FDG within the control volume yields the following governing equation

$$\phi \frac{\partial C_i}{\partial t} = \phi D \nabla^2 C_i - (1-\phi)(k_1 C_i - k_2 C_1), \tag{3}$$

which describes the 3D transport of [18F]FDG through the interstitium. The left-hand side of equation (3) denotes the local rate of change in the amount of [18F]FDG, the term $\phi D \nabla^2 C_i$ accounts for diffusive transport, and $(1-\phi)(k_1 C_i - k_2 C_1)$ represents the rate of exchange between cells and the interstitial fluid.

[18F]FDG, a glucose analog, undergoes facilitated transport across the vascular endothelium via glucose transporters (Chung et al., 2025). This process moves [18F]FDG across the vascular wall down its concentration gradient; thus, the diffusive flux ($J$) of [18F]FDG across the vascular endothelium can be expressed as $J = p_v(C_p - C_i)$, where $p_v$ is the permeability of the vascular walls to [18F]FDG (Figure 3(b)).

The diffusive flux of [18F]FDG in the mixture of cells and interstitial fluid is $-\phi D \nabla C_i$. To ensure continuity of [18F]FDG flux at the blood vessel boundary, we have the following boundary condition

$$-\phi D(\nabla C_i \cdot \boldsymbol{n}) = p_v(C_p - C_i) \text{ at the blood vessel boundary.} \tag{4}$$

Here, $\boldsymbol{n}$ is the unit normal vector to the vessel surface, directed from the plasma towards the interstitial fluid.

At the initial time, $t = 0$, the [18F]FDG concentrations are

$$C_i(t=0,\boldsymbol{x}) = C_1(t=0,\boldsymbol{x}) = C_2(t=0,\boldsymbol{x}) = 0. \tag{5}$$

We assume that the computational domain is symmetric at the boundaries. Thus, a zero-flux condition is applied for [18F]FDG transport at the domain boundary. Given that our domain dimensions (1 mm × 1 mm × 1 mm) are significantly larger than the typical intercapillary distance (50 μm), boundary effects are minor.

Furthermore, following intravenous injection, the input function representing [18F]FDG concentration in the plasma can be modeled as $C_p(t) = \Omega(t) \cdot I_d/V_d$ (Shiozaki et al., 2000), which is plotted in Figure 2(b) using the default parameters $I_d = 200$ MBq and $V_D = 10$ L, as listed in Table 1. Here, $I_d$ represents the injected dose (MBq), and $\Omega(t)$ is a standardized input function, which was obtained by averaging over 101 patients (Shiozaki et al., 2000). The initial volume of distribution of [18F]FDG is estimated as $V_d = 0.039 \cdot H^{0.8} \cdot W^{0.35}$, where $H$ and $W$ are the patient's height (cm) and weight (kg), respectively, and $V_d$ is in units of liter.

The radioactivity per unit volume of tissue is calculated as $C_r(t, \boldsymbol{x}) = [C_p(t)V_p + C_i(t, \boldsymbol{x})V_i + (C_1(t, \boldsymbol{x}) + C_2(t, \boldsymbol{x}))V_c]/V$. Its average over the 1 mm$^3$ spatial domain, denoted by $C_t(t)$, is what we can measure from PET imaging.

*2.1.2 Three-Tissue Compartment Model*

After moving from the plasma to the interstitial fluid via facilitated diffusion, [18F]FDG diffuses along concentration gradients, from regions near the vessels toward more distant regions, while simultaneously undergoing cellular uptake. This process is governed by three characteristic timescales: (1) the observation timescale ($T$), which represents the elapsed time between intravenous injection and [18F]FDG imaging, (2) the interstitial diffusion timescale ($l_{90\%}^2/D$), which is calculated based on the distance within which 90% of cells lie from their nearest vessels ($l_{90\%}$), and (3) the [18F]FDG uptake timescale ($k_1^{-1}$).

For static [18F]FDG-PET, the imaging timepoint $T$ is typically around 1 hour post-injection (Lin et al., 2005); but for dynamic [18F]FDG-PET, imaging begins within minutes of injection. If imaging happens at $T \ll l_{90\%}^2/D$, [18F]FDG will not have sufficient time to diffuse to cells distant from blood vessels. Consequently, the concentration in these distant regions will be significantly lower than in perivascular regions, resulting in spatial heterogeneity. Similarly, if $k_1^{-1} \ll l_{90\%}^2/D$ (i.e., cellular uptake is fast compared to the [18F]FDG diffusion), [18F]FDG is consumed rapidly as it diffuses away from the vasculature, resulting in heterogeneity as well.

Thus, we define a dimensionless parameter $\lambda_1 = D \cdot \min\{T, k_1^{-1}\}/l_{90\%}^2$. When $\lambda_1 \gg 1$ (i.e., $T \gg l_{90\%}^2/D$ and $k_1^{-1} \gg l_{90\%}^2/D$ are satisfied), the [18F]FDG concentration in the interstitial fluid will be approximately homogeneous. In this regime, the interstitium can be treated as a single lumped compartment; the total amount of [18F]FDG is approximated as $V_iC_i$, and the net inflow rate from the plasma is $p_vS(C_p - C_i)$. From equation (1), the net cellular uptake rate of [18F]FDG is $V_c(k_1C_i - k_2C_1)$. Applying the principle of mass conservation to the interstitial [18F]FDG yields the following governing equation:

$$V_i \frac{\mathrm{d}C_i}{\mathrm{d}t} = p_vS(C_p - C_i) - V_c(k_1C_i - k_2C_1), \tag{6}$$

where $S$ is the total surface area of the blood vessels. Equations (1), (2), and (6) constitute a three-tissue compartment model (Figure 4(a)), which differs from the used one reported in the literature (Figure 1 (b)) since the interstitial fluid volume ($V_i$) is not equal to the intracellular volume ($V_c$).

Note that equation (6) is consistent with equations (3) and (4). In a scenario where the [18F]FDG concentration in the interstitial fluid is approximately homogeneous, integrating equation (3) over the entire volume ($V_i + V_c$) consisting of the interstitium and cells yields

$$\int \phi \frac{\partial C_i}{\partial t} dv = \int (\phi D \nabla C_i \cdot \boldsymbol{n}') ds - \int (1 - \phi)(k_1C_i - k_2C_1)\, dv. \tag{7}$$

Here, $\boldsymbol{n}'$ is the unit normal vector to the vessel surface, directed from the interstitial fluid towards the plasma, i.e., $\boldsymbol{n}' = -\boldsymbol{n}$. Under the assumption of spatial homogeneity, the left-hand side of equation (7) becomes $V_i\,\mathrm{d}C_i/\mathrm{d}t$, and the cellular uptake term $\int(1-\phi)(k_1C_i-k_2C_1)\,\mathrm{dv}=V_c(k_1C_i-k_2C_1)$. The rate of inflow is $\int(\phi D\nabla C_i\cdot\boldsymbol{n}')\mathrm{ds}=\int p_v\big(C_p-C_i\big)\mathrm{ds}=p_vS(C_p-C_i)$. Thus, equation (6) can be rigorously derived from the 3D transport equations (3) and (4).

It is also worth noting that the diffusive term $\phi D\nabla^2C_i$ in equation (3) is non-zero, even though the [18F]FDG concentration in the interstitial fluid is approximately homogeneous. From the perspective of a three-tissue compartment model, once plasma [18F]FDG enters the interstitial fluid, its mass is assumed to distribute instantaneously throughout the entire interstitial compartment. Thus, for the control volume considered in equation (3), the diffusive inflow rate should be approximately equal to $\int\phi D\nabla^2C_i\mathrm{d}v=\phi D\nabla^2C_iv=p_vS(C_p-C_i)(v_i/V_i)$; thus, we have $\phi D\nabla^2C_i=p_vS(C_p-C_i)/(V_i+V_c)$.

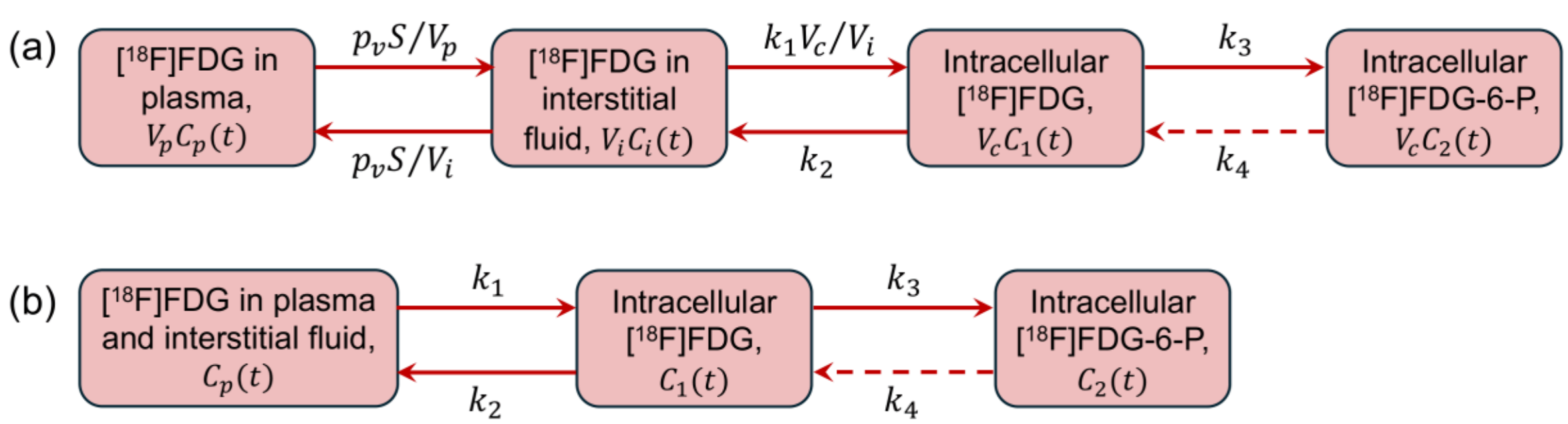

Figure 4. (a) Three-tissue compartment model, valid when $\lambda_1=D\cdot min\{T,k_1^{-1}\}/l_{90\%}^2\gg1$. (b) Two-tissue compartment model, valid when $\lambda_1=D\cdot min\{T,k_1^{-1}\}/l_{90\%}^2\gg1$ and $\lambda_2=(p_vS/V_c)\cdot min\{T,k_1^{-1}\}\gg1$.

*2.1.3 Two-Tissue Compartment Model*

Equation (6) can be rewritten as

$$\frac{V_i}{V_c}\cdot\frac{\mathrm{d}C_i}{\mathrm{d}\tilde{t}}-k_2TC_1=\frac{p_vST}{V_c}\left[C_p-C_i\left(1+\frac{k_1V_c}{p_vS}\right)\right],\tag{8}$$

where $\tilde{t}=t/T$. If $p_vST/V_c\gg1$ and $k_1V_c/(p_vS)\ll1$, then $C_i$ must approach $C_p$ to prevent divergence of the left-hand side of equation (7). In other words, if the conditions $V_c/(p_vS)\ll T$ and $V_c/(p_vS)\ll k_1^{-1}$ are satisfied, then $C_i\approx C_p$ and the three-tissue compartment model (Figure 4(a)) reduces to the widely used two-tissue compartment model (Figure 4(b)), where the plasma and interstitial fluid form a single compartment. Thus, we introduce the dimensionless parameter $\lambda_2=(p_vS/V_c)\cdot\min\{T,k_1^{-1}\}$ to quantify the vascular permeability relative to other processes. If $\lambda_2\gg1$, i.e., both conditions $V_c/(p_vS)\ll T$ and $V_c/(p_vS)\ll k_1^{-1}$ are satisfied, it is appropriate to represent the plasma and interstitial fluid as a single compartment.

From a physical perspective, the parameter $V_c/(p_vS)$ represents the characteristic timescale of facilitated diffusion across the vascular endothelium. If $V_c/(p_vS)$ and $l_{90\%}^2/D$ are both much smaller than the observation timescale ($T$) and the [18F]FDG uptake timescale ($k_1^{-1}$), then the vessels are sufficiently permeable that the difference in [18F]FDG concentration across the vessel walls is negligible, and [18F]FDG diffusion within the interstitium is sufficiently rapid. In this regime, [18F]FDG concentration in the interstitial fluid approximates that in the plasma, and the two-tissue compartment model assumption holds. Therefore, the two-tissue compartment model is valid when both $\lambda_1 \gg 1$ and $\lambda_2 \gg 1$.

## 2.2 Vasculature Dataset

To analyze [$^{18}$F]FDG transport from blood vessels to cells, we use a public dataset depicting the three-dimensional vasculature of the mouse inferior colliculus, a midbrain structure involved in auditory processing (Bumgarner and Nelson, 2022). We define our computational domain by selecting a subvolume composed of 1 million voxels ($100 \times 100 \times 100$), representing a total cubic volume of 1 mm$^3$, which is approximately the smallest volume resolvable by PET with current technology. Based on the voxel size $\Delta l$ of 10 μm, the blood vessels occupy 9.4% of the total volume, and the total vessel surface area ($S$) is 12 mm$^2$, which is calculated by multiplying $\Delta l^2$ by the number of surfaces that lie between vessel and cell voxels.

A histogram of the distance between each cell and its nearest vessels ($l_{\min}$) is shown in Figure 5. 90% of cells lie within 61 μm from the nearest vessels, a distance which we denote as $l_{90\%}$. This finding agrees with reports that human cells are generally less than 50-100 μm away from the nearest blood vessel (Baish et al., 2011).

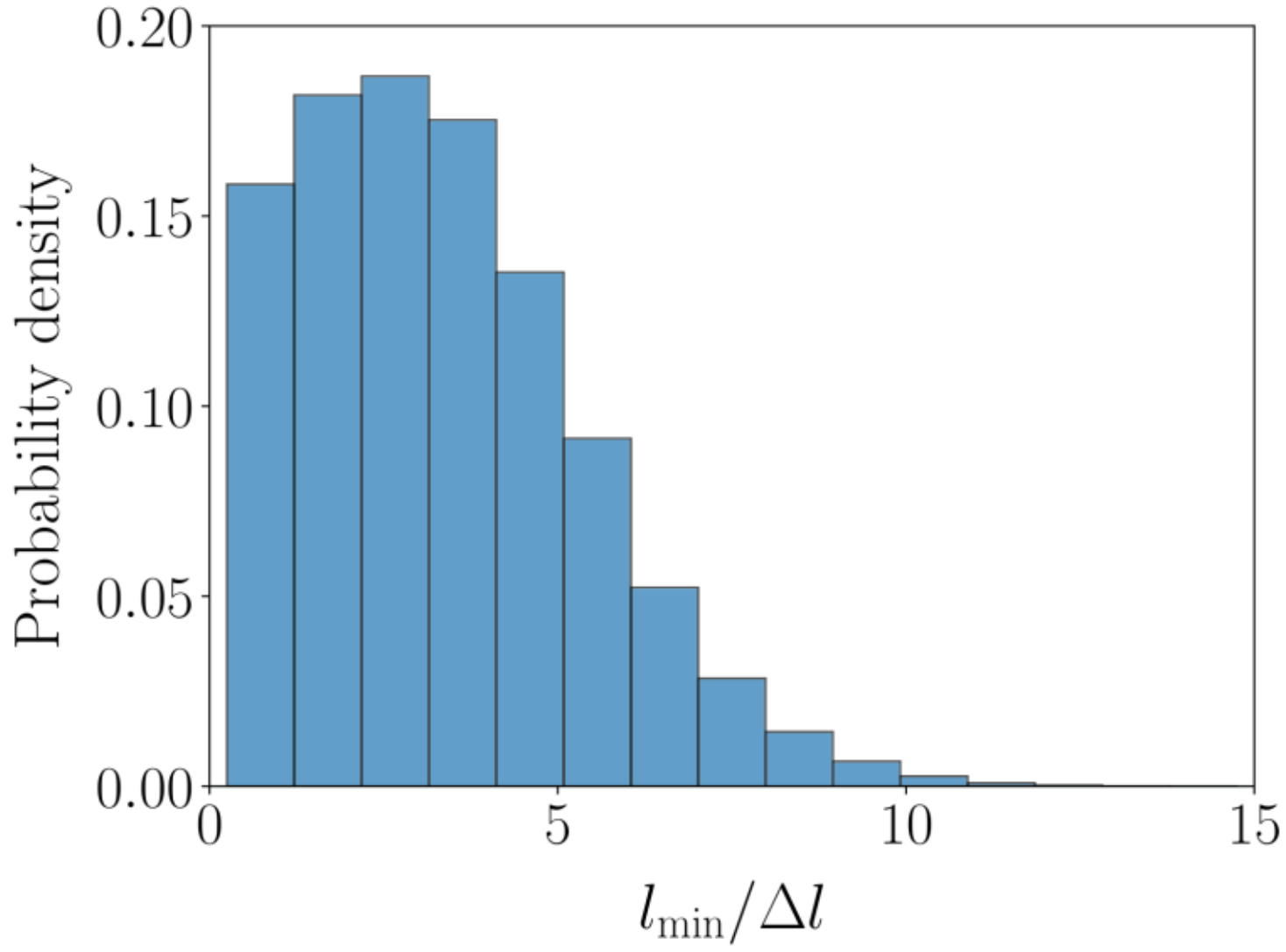

Figure 5. Distribution of distances between cells and their nearest blood vessels, measured in the inferior colliculus of a mouse.

## 2.3 Finite Difference Simulation Setup

To solve equations (1)-(5), we develop a finite difference solver in Python using the explicit Euler method with a time step size $\Delta t = \Delta l^2/6D$. The cubic computational mesh is aligned with the data voxel grid, and we have verified that the mesh resolution is sufficiently fine to achieve mesh-independent simulation results. The simulation takes approximately 30 minutes to run on Google Colab using the parameters listed in Table 1.

Unless otherwise specified, we use the parameter values listed in Table 1 in our computation. The intracellular fluid volume is set to 2.7 times the interstitial fluid volume (Watanabe, 2018), i.e., $V_c \approx 2.7V_i$. To compute the input function ($C_p$), the [$^{18}$F]FDG volume of distribution ($V_d$) is estimated assuming a patient height of $H = 170$ cm and a weight of $W = 60$ kg. The rate constants $k_1$, $k_2$, $k_3$, and $k_4$ are taken from the literature (Shiozaki et al., 2000). [18F]FDG diffusivity in the interstitium is assumed to be the same as glucose diffusivity and typically ranges from $10^{-11}$ m$^2$/s in multicellular tumor spheroids (Casciari et al., 1988) to $6 \times 10^{-10}$ m$^2$/s in cell culture medium (Suhaimi et al., 2015). The permeability of blood vessels to [18F]FDG depends on the expression of glucose transporters by endothelial cells and can vary significantly between different organs. A recent study (Chung et al., 2025) reported the blood-brain barrier permeability to [18F]FDG as 0.1 ml/min/cm$^3$. Given that the brain blood vessel diameters range from tens of micrometers to several millimeters, we estimate the permeability of brain vessels ($p_v$) to lie between $10^{-7}$ and $10^{-5}$ m/s. However, it is important to note that vascular permeability in tumors is often much higher than that of normal vessels (Maeda, 2012).

Table 1. Relevant physical parameters and their reference values used in our simulations.

| Physical parameters | Symbols | Values |
|---|---|---|
| [18F]FDG diffusivity in the interstitial fluid | $D$ | $6 \times 10^{-9}$ m$^2$/min |
| Permeability of [18F]FDG across blood vessels | $p_v$ | $6 \times 10^{-6}$ m/min |
| Rate constant for [18F]FDG influx into cells | $k_1$ | 0.102 min$^{-1}$ |
| Rate constant for [18F]FDG efflux to the interstitial fluid | $k_2$ | 0.13 min$^{-1}$ |
| Phosphorylation rate of [18F]FDG | $k_3$ | 0.062 min$^{-1}$ |
| Dephosphorylation rate of [18F]FDG-6-P | $k_4$ | 0.0068 min$^{-1}$ |
| Voxel length | $\Delta l$ | 10 µm |
| Injected dose | $I_d$ | 200 MBq |
| Volume of distribution of [18F]FDG | $V_d$ | 10 L |
| Volume ratio of intracellular fluid to interstitial fluid | $V_c/V_i$ | 2.7 |

## 3. Results

### 3.1 Impact of Vascular Permeability, [18F]FDG Diffusivity, and Uptake Rate Constant

The uptake of [18F]FDG by tumors, as measured by PET, is commonly understood as a measure of the intrinsic metabolic activity of the cancer cells, reflecting their elevated expression of glucose transporters and hexokinase enzyme. However, other parameters such as vascular permeability, vascular architecture, and extracellular [18F]FDG diffusivity also contribute significantly to the measured PET signal.

Compartment models assume homogeneous distribution of [18F]FDG within the interstitial fluid. In this subsection, we examine this assumption by analyzing the spatial homogeneity of radioactivity per unit volume of tissue, $C_r(t, x)$, under different scenarios. Additionally, we compare the simulated values of $C_t(t)$, corresponding to voxel intensity measured by PET, for the different modeling strategies described in Section 2.

Note that the value of $\lambda_1$ reflects the level of spatial homogeneity of [18F]FDG within the interstitial fluid and, therefore, whether the three-tissue compartment model is a suitable representation of the uptake process. Likewise, the value of $\lambda_2$ characterizes vascular permeability and can be used to determine whether the three-tissue compartment model can be reduced to the simpler two-tissue compartment model.

Using the parameter values listed in Table 1, we compute the temporal evolution of the PET signal $C_t(t)$ and the microscopic spatial distribution of the radioactivity $C_r(t, \boldsymbol{x})$ at 1, 15, and 60 minutes (Figure 5(a)). In this scenario, $\lambda_1 = 16$ is much greater than 1, indicating that FDG distribution at $t = 1$ hour is approximately homogeneous, which explains the agreement between the finite-difference simulation, which accounts for diffusion, and the three-tissue compartment model. The value of $\lambda_2 = 1.1$ suggests limited vascular permeability, so we observe a discrepancy between two- and three-tissue compartment models.

Note that for early time points (e.g. $t < 3$ min), $\lambda_1$ is smaller than 1, indicating heterogeneous interstitial [$^{18}$F]FDG concentration. However, the absolute [$^{18}$F]FDG concentration in the interstitial fluid remains low during the earliest time points, and hence the spatial heterogeneity has minimal impact on the flux of [$^{18}$F]FDG from blood vessels to the interstitial fluid. This explains the minor discrepancy between the three-tissue compartment model and the finite-difference simulation results at early time points in Figure 6(a).

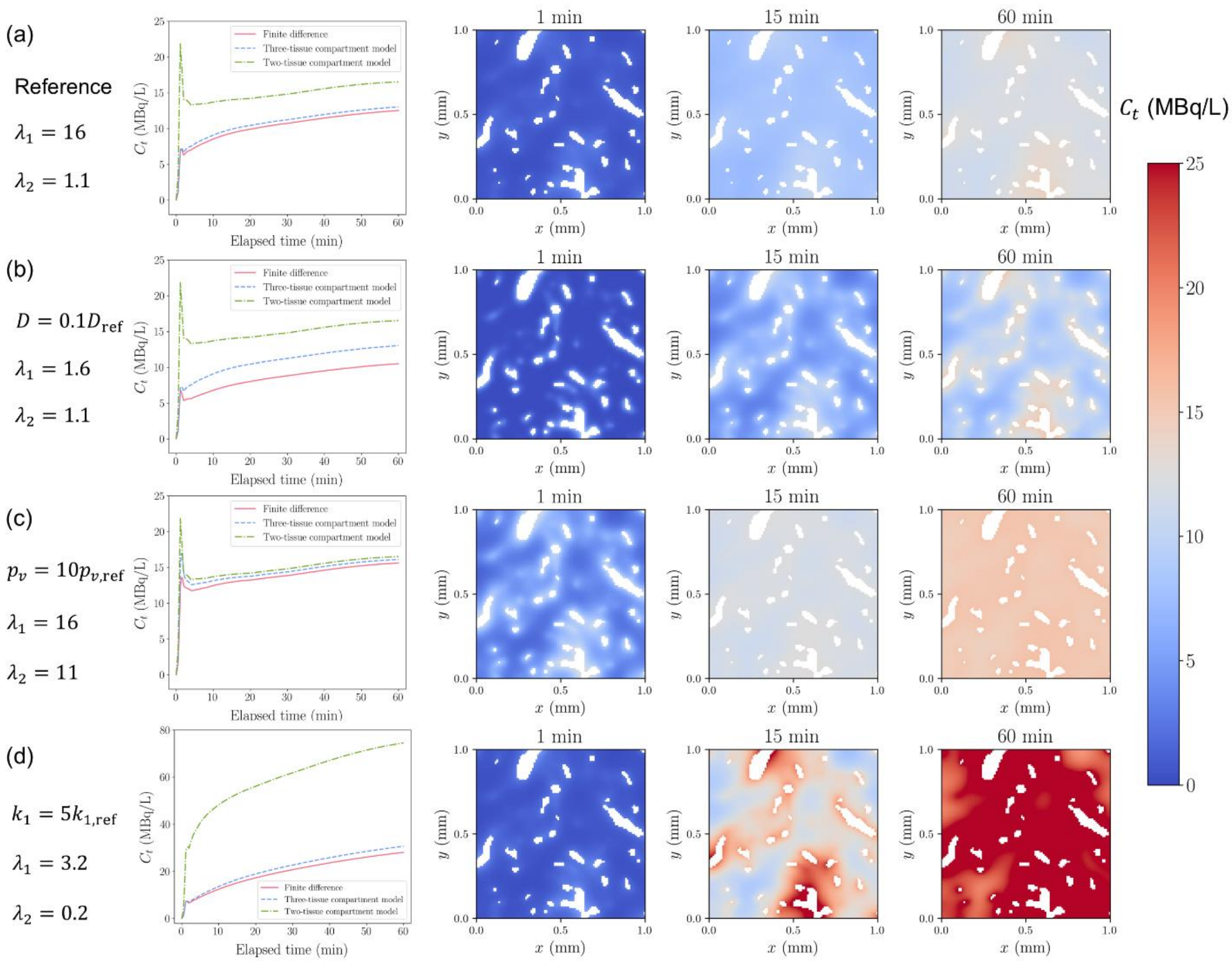

Figure 6. Temporal evolution of the simulated PET signal $C_t(t)$ and microscopic distribution of the radioactivity $C_r(t, \boldsymbol{x})$, under various conditions: (a) nominal parameter values listed in Table 1; (b) reduced FDG diffusivity, $D = 10^{-11}$ m$^2$/s; (c) increased vascular permeability, $p_v = 10^{-6}$ m/s; and (d) increased rate constant for [$^{18}$F]FDG influx into cells, $k_1 = 0.51$ min$^{-1}$. A 2D section of the mid-plane is used here for illustration. The vessels are marked in white.

We take the scenario shown in Figure 6(a) as the reference case, denoting its parameter set with the subscript "ref", and then reduce the interstitial [$^{18}$F]FDG diffusivity ($D$) tenfold. In this scenario, $\lambda_1$ decreases to 1.6, and we observe a more heterogenous [$^{18}$F]FDG distribution at $t = 1$ hour (Figure 6(b)). Because of this, the three-tissue compartment model result deviates significantly from the accurate finite difference simulation.

Likewise, when vascular permeability is increased tenfold ($\lambda_1 = 16$ and $\lambda_2 = 11$), [$^{18}$F]FDG distribution is homogeneous, and both the three-tissue compartment model and the finite-difference simulation converge to the two-tissue compartment model (Figure 6(c)). By contrast, when the [$^{18}$F]FDG uptake rate constant ($k_1$) is increased fivefold, $\lambda_2$ decreases to 0.2, leading to a pronounced discrepancy between the two- and three- tissue compartment models (Figure 6(d)).

To further enhance our confidence in using $\lambda_1$ and $\lambda_2$ to characterize the validity of the two- and three-tissue compartment models, we individually vary [$^{18}$F]FDG diffusivity, uptake rate constant, and vascular permeability, and then compare the simulated values of $C_t(t = 1\,\mathrm{h})$ obtained with different modeling strategies (Figure 7). We observe that the three-tissue compartment model (Figure 4(a)) agrees with the accurate finite-difference simulation when $\lambda_1 \gg 1$, and both converge to the two-tissue compartment model (Figure 4(b)) when $\lambda_2 \gg 1$ is also satisfied.

Based on the above analyses, the impacts of vascular permeability, vascular architecture, cellular uptake, [$^{18}$F]FDG diffusivity, and acquisition time are combined into two dimensionless parameters, $\lambda_1$ and $\lambda_2$. The values of these parameters determine the validity of the two- and three-tissue compartment models, as defined in Figure 4. More importantly, the definition and interpretation of the rate constants is consistent across the three modeling strategies.

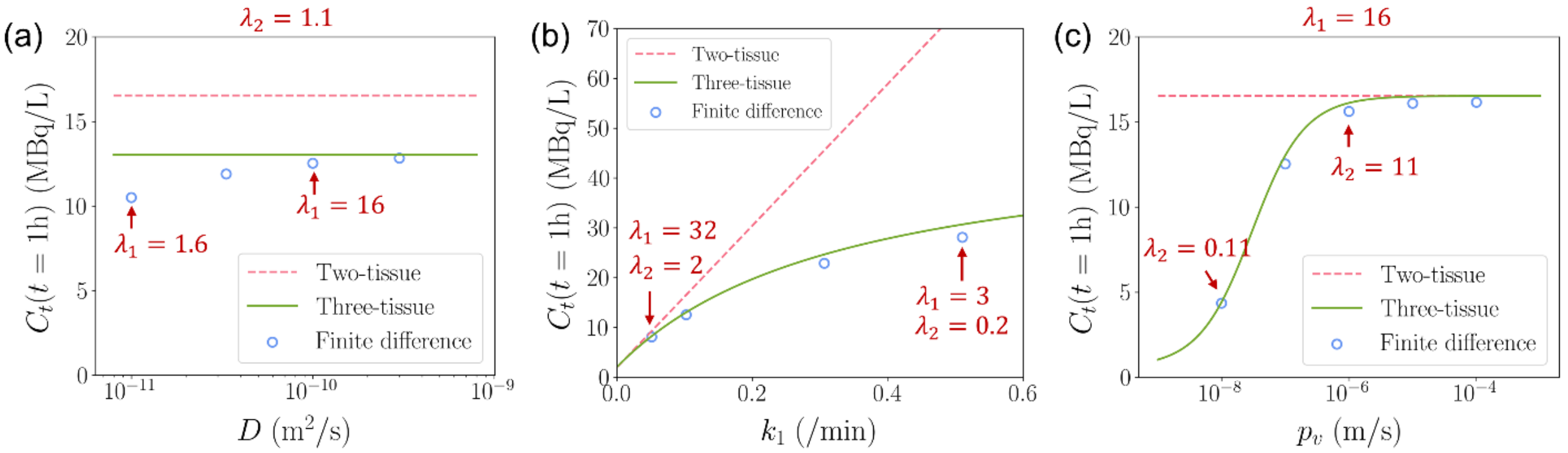

Figure 7. Simulations of the measured PET signal $C_t(t = 1\,h)$ for different [$^{18}$F]FDG diffusivities, uptake rate constants, and vessel permeabilities. Red dashed line: two-tissue compartment model (Figure 4(b)); Green solid line: three-tissue compartment model (Figure 4(a)); Blue circles: Accurate finite difference simulation.

## 3.2 Impact of Vascular Architecture

Compared to healthy tissues, tumors typically have lower [$^{18}$F]FDG diffusivity, higher cellular metabolism, and irregular vascular architecture. To investigate how these factors jointly affect [$^{18}$F]FDG distribution in the interstitial fluid and the average radioactivity per unit volume of tissue, $C_t(t)$, we first consider a tissue with reduced [$^{18}$F]FDG diffusivity ($D = 10^{-11}$ m$^2$/s) and elevated uptake ($k_1 = 0.51$ min$^{-1}$) relative to the reference case. Under these conditions, the spatial distribution of $C_r(t, x)$ (Figure 8(a)) becomes significantly more heterogeneous than in the reference scenario (Figure 6(a)), and the three-tissue compartment model prediction, $C_t(t = 1\,\mathrm{h}) = 30.7$ MBq/L, deviates by 46% from the more accurate finite difference simulation result, $C_t(t = 1\,\mathrm{h}) = 21.0$ MBq/L (Figure 9).

Next, we investigate the impact of vascular architecture by removing different numbers of vessels from the dataset. By segmenting vessels using 6-connectivity, where voxels are considered connected if they share a face, we identify 253 vessels within the computational domain. We then randomly remove 40% and 70% of the labeled vessels. As a result, the total vessel surface area ($S$) decreases while the distance within which 90% of cells lie from their nearest vessels ($l_{90\%}$) increases from 61 μm to 119 μm and 185 μm, respectively. When 70% of vessels are removed, the discrepancy between the three-tissue compartment model and the diffusion-incorporated simulation increases to 90% (Figure 9).

These results indicate that under conditions of low diffusivity, high cellular metabolism, and sparse vascular architecture, [$^{18}$F]FDG distribution is highly heterogeneous, and the use of two- and three-tissue compartment models to interpret PET images will underestimate the cellular uptake rate that reflects the underlying metabolic activity of the cells.

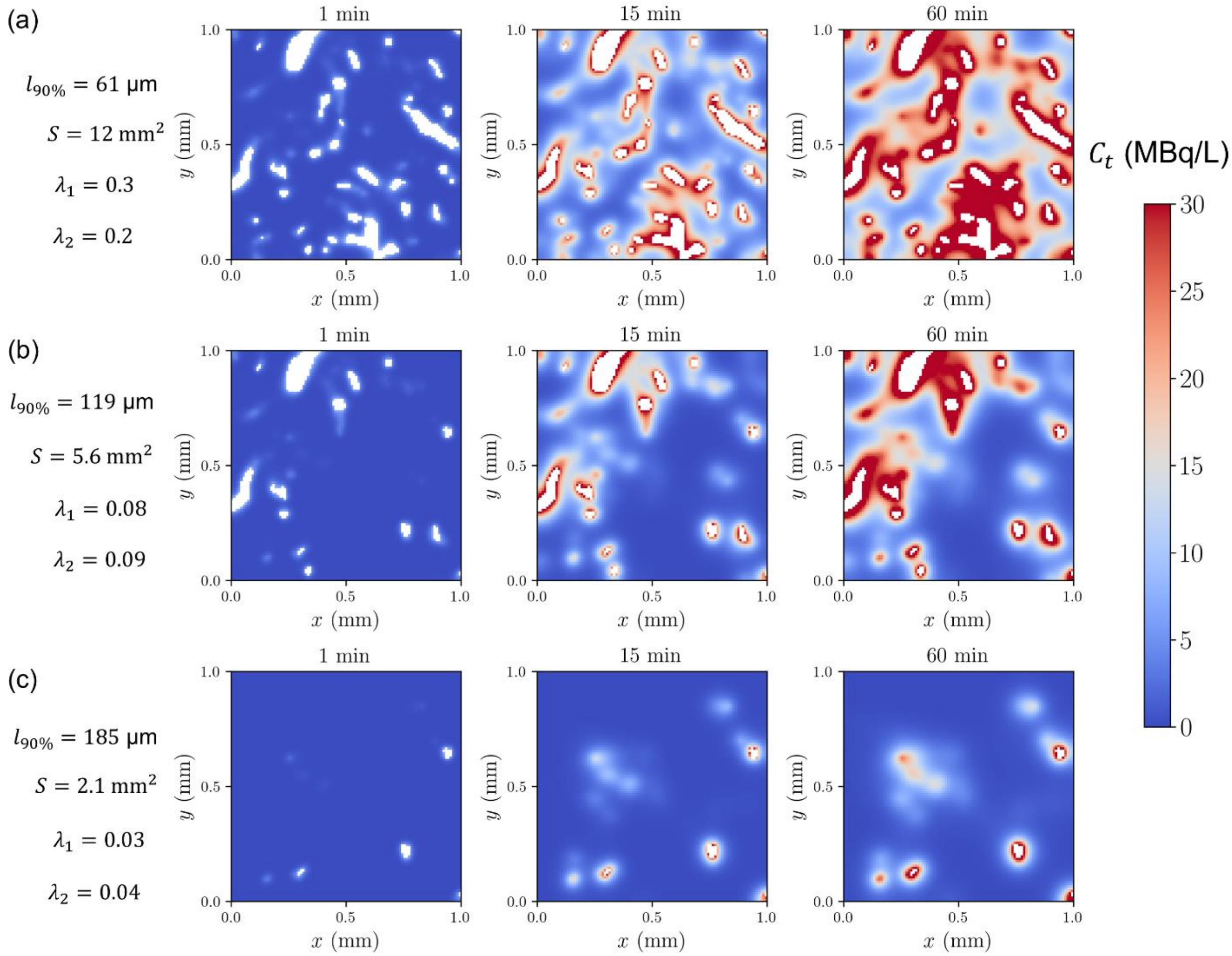

Figure 8. Microscopic Spatiotemporal distribution of the radioactivity per unit volume of tissue, $C_T(t, \boldsymbol{x})$, for varying vascular architectures. Reference parameter values are used, except for $D = 10^{-11}$ m$^2$/s and $k_1 = 0.51$ min$^{-1}$. (a) Default vascular architecture with $l_{90\%} = 61$ μm; (b) 40% of vessels are removed, and $l_{90\%} = 119$ μm; (c) 70% of vessels are removed, and $l_{90\%} = 185$ μm.

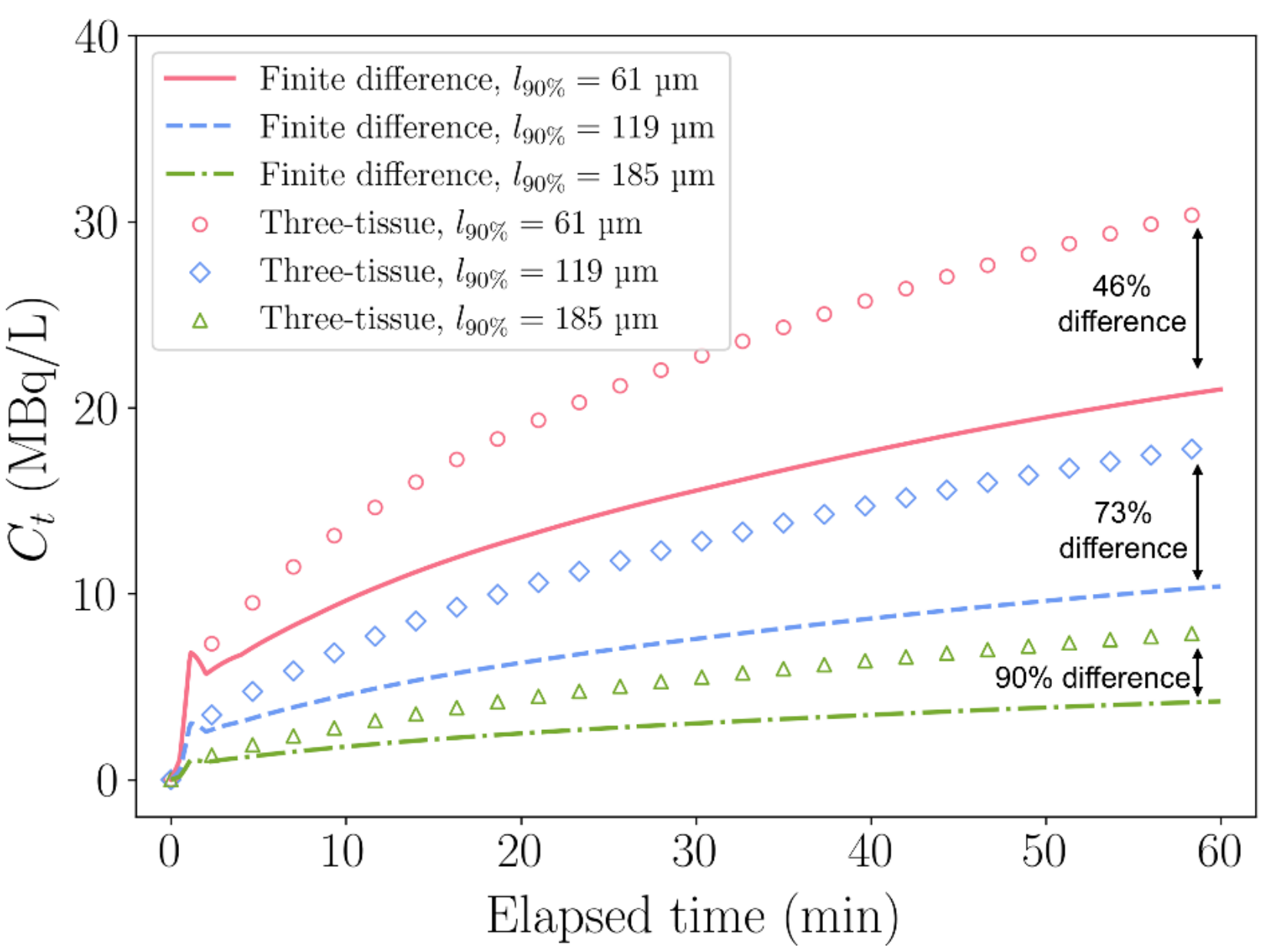

Figure 9. Temporal evolution of the PET signal $C_t(t)$ with $D = 10^{-11}$ m²/s and $k_1 = 0.51$ min⁻¹. Red solid, blue dashed, and green dash-dotted lines are the result of the accurate finite difference simulation with $l_{90\%} = 61$ μm, 119 μm, and 185 μm, respectively. Red circles, blue squares, and green triangles are obtained using the three-tissue compartment model with $l_{90\%} = 61$ μm, 119 μm, and 185 μm, respectively.

### 3.3 Estimation of Cellular Metabolic Rates

The two-tissue compartment model is commonly used to analyze dynamic [$^{18}$F]FDG-PET data to estimate cellular metabolic rates ($k_1$, $k_2$, $k_3$, and $k_4$). To evaluate its prediction accuracy, we fit the two-tissue compartment model (Figure 4(b)) to time-activity curves $C_t(t)$ generated using finite-difference model and the nominal values for $k_1$, $k_2$, $k_3$, and $k_4$ listed in Table 1 under different $\lambda_1$ and $\lambda_2$ conditions. We use the Python package pymcmcstat (Zhong et al., 2025) to perform Bayesian model calibration and use the mean-squared error of the predicted $C_t(t)$ as the loss function. The search domain is $0 \leq k_1 \leq 1$ min⁻¹, $0 \leq k_2 \leq 1$ min-1, $0 \leq k_3 \leq 0.2$ min⁻¹, and $0 \leq k_4 \leq 0.2$ min⁻¹. Considering the search may be trapped in local minima, we start the search using ten randomly generated sets of initial values.

We consider the following two scenarios: (1) $D = 10^{-10}$ m²/s and $p_v = 10^{-4}$ m/s (i.e., $\lambda_1 = 16$ and $\lambda_2 = 1100$), a case where the two-tissue compartment model is valid; and (2) $D = 10^{-11}$ m²/s and $p_v = 10^{-7}$ m/s (i.e., $\lambda_1 = 1.6$ and $\lambda_2 = 1.1$), a case where there is a significant discrepancy between the two-tissue compartment model and the finite difference simulation, as shown in Figure 6(b). We observe that, in both scenarios, the best-fit two-tissue compartment model captures the long-term behavior of $C_t(t)$, as shown in Figure 10. However, the fitted cellular uptake rate $k_1$ is smaller than its nominal value ($k_1 = 0.102$ min⁻¹). In the first scenario ($\lambda_1 = 16$ and $\lambda_2 = 1100$), the fitted $k_1 = 0.085$ min⁻¹ is 15% smaller than the nominal value. In the second scenario ($\lambda_1 = 1.6$ and $\lambda_2 = 1.1$), the fitted $k_1 = 0.02$ min⁻¹ is only one-sixth of the nominal value. Therefore, the two-tissue compartment model underestimates the cellular uptake rate, particularly when $\lambda_1$ and $\lambda_2$ are not much greater than 1 (i.e., low [$^{18}$F]FDG diffusivity and limited vascular permeability). Hence, the estimated rate constant reflects not only the intrinsic metabolic properties of the cells, but also the delivery of the radiotracer through the vasculature and interstitial space.

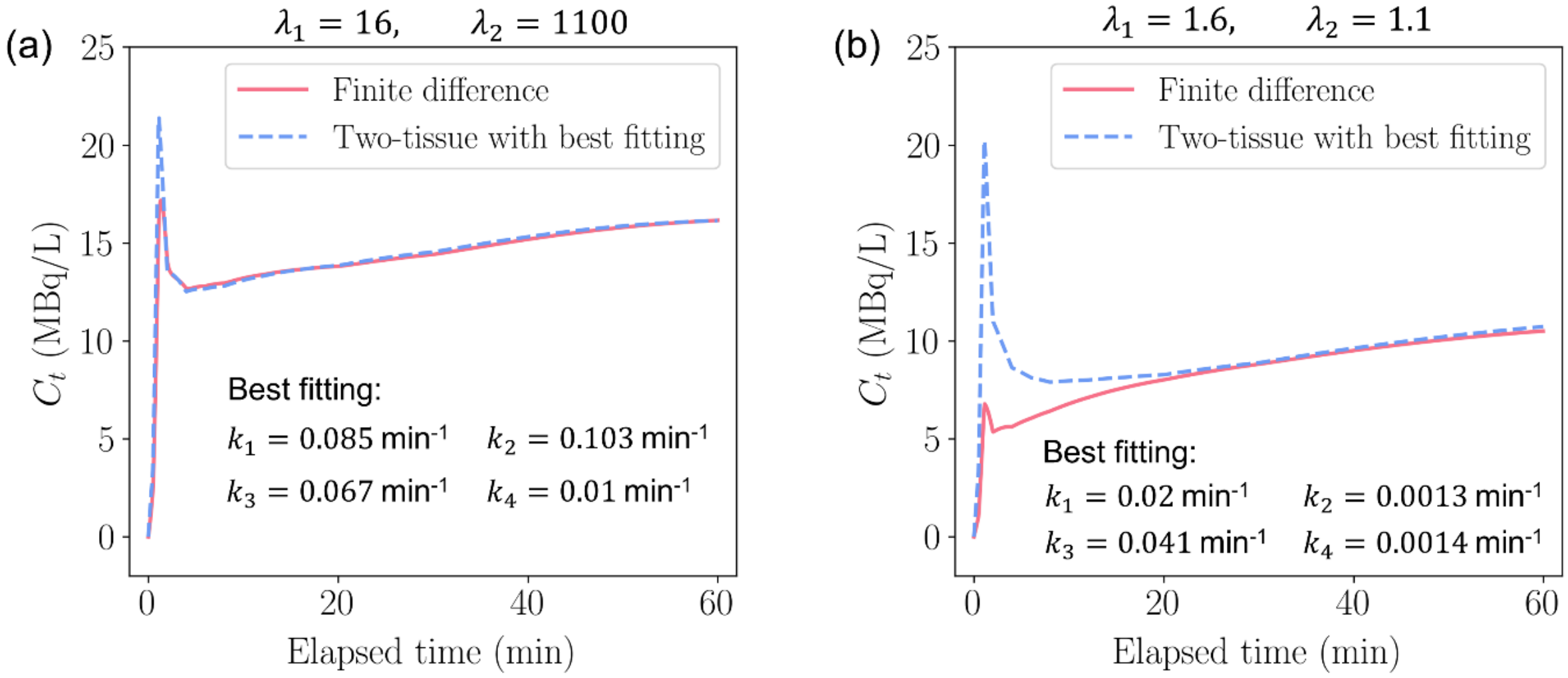

Figure 10. Comparison between the simulated finite difference data and the best-fit two-tissue compartment model. The simulation data is generated using nominal values $k_1 = 0.102$ min$^{-1}$, $k_2 = 0.13$ min$^{-1}$, $k_3 = 0.062$ min$^{-1}$, and $k_4 = 0.0068$ min$^{-1}$. (a) $\lambda_1 = 16$ and $\lambda_2 = 1100$; (b) $\lambda_1 = 1.6$ and $\lambda_2 = 1.1$. The fitted rate constants are indicated on each figure.

## 4. Discussion and Conclusion

In this study, we investigate the influence of vascular architecture, vascular permeability, [$^{18}$F]FDG diffusivity, and cellular uptake rate constant on the spatial distribution of [$^{18}$F]FDG. We introduce three modeling strategies that are consistent in their definitions and interpretations of parameters and rate constants. To select the appropriate model, we propose two dimensionless parameters, $\lambda_1 = D \cdot \min\{T, k_1^{-1}\}/l_{90\%}^2$ and $\lambda_2 = (p_v S/V_c) \cdot \min\{T, k_1^{-1}\}$. Our results show that when $\lambda_1 \gg 1$, FDG distribution can be considered homogeneous, and the three-tissue compartment model (Figure 4(a)) can be safely used. If $\lambda_2 \gg 1$ is further satisfied, then vascular permeability is sufficiently high for the plasma and interstitial fluid to be treated as a single compartment; in this case, this three-tissue compartment model reduces to the classical two-tissue compartment model (Figure 4(b)).

The applicability of the two-tissue and three-tissue compartment models are time-dependent, as both $\lambda_1$ and $\lambda_2$ are proportional to $\min\{T, k_1^{-1}\}$. At early time points, $\lambda_1$ and $\lambda_2$ are smaller than 1, indicating both three-tissue and two-tissue compartment models are inaccurate. However, the significance of this initial inaccuracy depends on its cumulative effect over time. The homogeneity assumption introduces only minor error at early stages because the interstitial [$^{18}$F]FDG concentration is much lower than that in the plasma, minimizing the impact of heterogeneity on [$^{18}$F]FDG flux from plasma to the interstitial fluid. Consequently, if $\lambda_1$ rapidly increases to exceed 1, the cumulative error remains limited, ensuring the predictive reliability of the three-tissue compartment model at later stages.

In contrast, the two-tissue compartment model depicted in Figure 4(b), relies on the stricter assumption that the interstitial [$^{18}$F]FDG concentration is equal to that in the plasma. For this model to be valid, both $\lambda_1$ and $\lambda_2$ must quickly increase and exceed 1, ensuring conditions of both spatial homogeneity and high vascular permeability are met.

The parameter $\lambda_1$ is inversely proportional to $l_{90\%}^2$, and $\lambda_2$ is proportional to $S/V_c$. Compared to humans, small animals such as mice have shorter intercapillary distances and smaller vessel diameters (i.e., larger $S/V_c$) (Müller et al., 2008; Todorov et al., 2020). Therefore, the values of $\lambda_1$ and $\lambda_2$ in mice are expected to be higher than those in human, assuming similar vascular permeability and cellular uptake rate constants. This implies that while compartment models may appear valid in mice, their direct application to humans can be misleading and potentially result in misinterpretation of PET data.

Aggressive tumors often have significantly high vascular permeability, low [$^{18}$F]FDG diffusivity, elevated cellular metabolism, and irregular vascular architecture (West et al., 2001). In regions with dense vessel networks, corresponding to short vessel-cell distances and large vessel surface areas, the values of $\lambda_1$ and $\lambda_2$ may exceed 1, resulting in reasonable predictions by compartment models. By contrast, in regions with sparse vessel networks, compartment models are prone to substantial errors, and their use could underestimate cellular metabolic rates.

It is important to note that the three-tissue compartment model in Figure 4(a) has a different structure and rate constants compared to the model shown in Figure 1(b). This difference highlights an inconsistency between the two- and three-tissue compartment models depicted in Figure 1. Therefore, to ensure consistent interpretation of rate constants, the modeling strategies presented in Section 2 should be used for PET image analysis. Additionally, the dimensionless parameters $\lambda_1$ and $\lambda_2$ quantitatively integrate the effect of vascular architecture, [$^{18}$F]FDG transport dynamics, and cellular uptake kinetics, providing a practical way to delineate the scenarios where each compartment model is applicable.

There are several limitations to our modeling approach. First, we neglect convective transport, which may be appropriate for small molecules such as [$^{18}$F]FDG, depending on local [$^{18}$F]FDG diffusivity and interstitial fluid velocity (Dewhirst and Secomb, 2017). Second, our model ignores lymphatic drainage, an assumption that may hold for tumor tissues where lymphatic vessels are absent (Baxter and Jain, 1989; Dewhirst and Secomb, 2017). In tissues with lymphatic drainage, [$^{18}$F]FDG elimination should be accounted for by either adding a sink term to the right-hand side of equation (3) or modeling drainage at lymphatic walls if their architecture is known. Third, we assume a constant cell packing fraction and rate constants within each PET voxel. Fourth, the partial-volume effects, positron-range blurring, limited count statistics, and image reconstruction losses bring uncertainties to the measured PET signal. This uncertainty can compromise the accuracy of estimated cellular metabolic rates, which needs further investigation.

This work not only provides guidance for selecting the appropriate model in [$^{18}$F]FDG-PET analysis but also demonstrates that voxel intensity in PET images can be influenced by vascular architecture and physiological parameters under certain conditions, thus only partially reflecting the underlying metabolic rate constants. It also highlights the importance of considering vascular architecture and physiological parameters when translating [$^{18}$F]FDG-PET analyses of cellular metabolism from animal models to humans.

## 5. Acknowledgements

Research reported in this publication was supported by the National Institute of Biomedical Imaging and Bioengineering of the National Institutes of Health under Awards No. R01EB030367,

R01CA268514, and R21EB034534. The content is solely the responsibility of the authors and does not necessarily represent the official views of the National Institutes of Health.

**6. References**

Baish, J.W., Stylianopoulos, T., Lanning, R.M., Kamoun, W.S., Fukumura, D., Munn, L.L., Jain, R.K., 2011. Scaling rules for diffusive drug delivery in tumor and normal tissues. Proc. Natl. Acad. Sci. 108, 1799–1803. https://doi.org/10.1073/pnas.1018154108

Baxter, L.T., Jain, R.K., 1989. Transport of fluid and macromolecules in tumors. I. Role of interstitial pressure and convection. Microvasc. Res. 37, 77–104. https://doi.org/10.1016/0026-2862(89)90074-5

Bertoldo, A., Peltoniemi, P., Oikonen, V., Knuuti, J., Nuutila, P., Cobelli, C., 2001. Kinetic modeling of [ $^{18}$ F]FDG in skeletal muscle by PET: a four-compartment five-rate-constant model. Am. J. Physiol.-Endocrinol. Metab. 281, E524–E536. https://doi.org/10.1152/ajpendo.2001.281.3.E524

Bumgarner, J.R., Nelson, R.J., 2022. Open-source analysis and visualization of segmented vasculature datasets with VesselVio. Cell Rep. Methods 2, 100189. https://doi.org/10.1016/j.crmeth.2022.100189

Casciari, J.J., Sotirchos, S.V., Sutherland, R.M., 1988. Glucose Diffusivity in Multicellular Tumor Spheroids. Cancer Res. 48, 3905–3909.

Chung, K.J., Abdelhafez, Y.G., Spencer, B.A., Jones, T., Tran, Q., Nardo, L., Chen, M.S., Sarkar, S., Medici, V., Lyo, V., Badawi, R.D., Cherry, S.R., Wang, G., 2025. Quantitative PET imaging and modeling of molecular blood-brain barrier permeability. Nat. Commun. 16. https://doi.org/10.1038/s41467-025-58356-7

Dewhirst, M.W., Secomb, T.W., 2017. Transport of drugs from blood vessels to tumour tissue. Nat. Rev. Cancer 17, 738–750. https://doi.org/10.1038/nrc.2017.93

Dimitrakopoulou-Strauss, A., Pan, L., Sachpekidis, C., 2021. Kinetic modeling and parametric imaging with dynamic PET for oncological applications: general considerations, current clinical applications, and future perspectives. Eur. J. Nucl. Med. Mol. Imaging 48, 21–39. https://doi.org/10.1007/s00259-020-04843-6

Kashkooli, F.M., Abazari, M.A., Soltani, M., Ghazani, M.A., Rahmim, A., 2022. A spatiotemporal multi-scale computational model for FDG PET imaging at different stages of tumor growth and angiogenesis. Sci. Rep. 12, 10062. https://doi.org/10.1038/s41598-022-13345-4

Kotasidis, F.A., Tsoumpas, C., Rahmim, A., 2014. Advanced kinetic modelling strategies: towards adoption in clinical PET imaging. Clin. Transl. Imaging 2, 219–237. https://doi.org/10.1007/s40336-014-0069-8

Lin, W.-Y., Tsai, S.-C., Hung, G.-U., 2005. Value of delayed 18F-FDG-PET imaging in the detection of hepatocellular carcinoma: Nucl. Med. Commun. 26, 315–321. https://doi.org/10.1097/00006231-200504000-00003

Maeda, H., 2012. Vascular permeability in cancer and infection as related to macromolecular drug delivery, with emphasis on the EPR effect for tumor-selective drug targeting. Proc. Jpn. Acad. Ser. B 88, 53–71. https://doi.org/10.2183/pjab.88.53

Morris, E.D., Endres, C.J., Schmidt, K.C., Christian, B.T., Jr, R.F.M., Fisher, R.E., 2004. Kinetic Modeling in Positron Emission Tomography, in: Emission Tomography: The Fundamentals of PET and SPECT, 1. Academic Press, San Diego, CA, pp. 499–540.

Müller, B., Lang, S., Dominietto, M., Rudin, M., Schulz, G., Deyhle, H., Germann, M., Pfeiffer, F., David, C., Weitkamp, T., 2008. High-resolution tomographic imaging of microvessels, in: Developments in X-Ray Tomography VI. Presented at the Proc. of SPIE, San Diego, California, United States. https://doi.org/10.1117/12.794157

Oleschko, K., 1998. Delesse principle and statistical fractal sets: 1. Dimensional equivalents. Soil Tillage Res. 49, 255–266. https://doi.org/10.1016/S0167-1987(98)00179-2

Pantel, A.R., Viswanath, V., Muzi, M., Doot, R.K., Mankoff, D.A., 2022. Principles of Tracer Kinetic Analysis in Oncology, Part II: Examples and Future Directions. J. Nucl. Med. 63, 514–521. https://doi.org/10.2967/jnumed.121.263519

Quon, A., Gambhir, S.S., 2005. FDG-PET and Beyond: Molecular Breast Cancer Imaging. J. Clin. Oncol. 23, 1664–1673. https://doi.org/10.1200/JCO.2005.11.024

Shahvandi, M.K., Soltani, M., Kashkooli, F.M., Saboury, B., Rahmim, A., 2022. Spatiotemporal multi-scale modeling of radiopharmaceutical distributions in vascularized solid tumors. Sci. Rep. 12, 14582.

Shiozaki, T., Sadato, N., Senda, M., Ishii, K., Tsuchida, T., Yonekura, Y., Fukuda, H., Konishi, J., 2000. Noninvasive Estimation of FDG Input Function for Quantification of Cerebral Metabolic Rate of Glucose: Optimization and Multicenter Evaluation. J. Nucl. Med. 41, 1612–1618.

Suhaimi, H., Wang, S., Das, D.B., 2015. Glucose diffusivity in cell culture medium. Chem. Eng. J. 269, 323–327. https://doi.org/10.1016/j.cej.2015.01.130

Todorov, M.I., Paetzold, J.C., Schoppe, O., Tetteh, G., Shit, S., Efremov, V., Todorov-Völgyi, K., Düring, M., Dichgans, M., Piraud, M., Menze, B., Ertürk, A., 2020. Machine learning analysis of whole mouse brain vasculature. Nat. Methods 17, 442–449. https://doi.org/10.1038/s41592-020-0792-1

Watanabe, T., 2018. Biophysical basis of physiology and calcium signaling mechanism in cardiac and smooth muscle. Academic Press.

West, C., Cooper, R., Loncaster, J., Wilks, D., Bromley, M., 2001. Tumor Vascularity: A Histological Measure of Angiogenesis and Hypoxia. Cancer Res. 61, 2907–2910.

Zhong, X., Nguyen, H.T.M., Takematsu, E., Pratx, G., 2025. Diffusion-aware compartment model of the cellular uptake of 18F-fluorodeoxyglucose. Phys. Rev. E 111, 044409.

Zhu, A., Lee, D., Shim, H., 2011. Metabolic Positron Emission Tomography Imaging in Cancer Detection and Therapy Response. Semin. Oncol. 38, 55–69. https://doi.org/10.1053/j.seminoncol.2010.11.012